\documentclass[aps,prd,reprint,superscriptaddress,amsmath,amssymb,floatfix]{revtex4-1}

\usepackage{graphicx}
\usepackage{color}
\usepackage{dcolumn}
\usepackage{bm}
\usepackage{slashed}
\usepackage{placeins}

\begin{document}

\title{Resonant bound orbits and kludge waveforms in rotating Konoplya-Zhidenko black hole spacetime}

\author{Qiyu Dai}
\affiliation{Department of Physics, Key Laboratory of Low Dimensional Quantum Structures and Quantum Control of Ministry of Education, and Synergetic Innovation Center for Quantum Effects and Applications, Hunan Normal University, Changsha, Hunan 410081, P. R. China}

\author{Xiongjun Fang}
\email[Corresponding author: ]{fangxj@hunnu.edu.cn}
\affiliation{Department of Physics, Key Laboratory of Low Dimensional Quantum Structures and Quantum Control of Ministry of Education, and Synergetic Innovation Center for Quantum Effects and Applications, Hunan Normal University, Changsha, Hunan 410081, P. R. China}

\author{Xiao-Mei Kuang}
\email[]{xmeikuang@yzu.edu.cn}
\affiliation{ Center for Gravitation and Cosmology, College of Physical Science and Technology, Yangzhou University, Yangzhou, Jiangsu 225009, P. R. China}

\author{Jiliang Jing}
\affiliation{Department of Physics, Key Laboratory of Low Dimensional Quantum Structures and Quantum Control of Ministry of Education, and Synergetic Innovation Center for Quantum Effects and Applications, Hunan Normal University, Changsha, Hunan 410081, P. R. China}

\begin{abstract}

We investigate timelike bound motion, resonant periodic orbits, and their gravitational-wave signatures in the rotating Konoplya–Zhidenko (KZ) black hole spacetime. Using the separability of the Hamilton–Jacobi equation, we parameterize eccentric and inclined bound orbits by $(p,e,z_1)$, derive the corresponding constants of motion $(E,L_z,Q)$, and use the orbital frequencies to identify resonant configurations. We study a range of resonances, including radial–polar resonances of inclined orbits and radial–azimuthal resonances of equatorial eccentric orbits. We further construct physically scaled quadrupole-kludge waveforms for representative equatorial resonant orbits and analyze their frequency-domain characteristics. Our results show that the KZ deformation shifts the resonance locations and modifies both the orbital trajectories and the resulting gravitational-wave signals. The corresponding characteristic strain lies predominantly in the millihertz band, placing these signals in the frequency range relevant to space-based gravitational wave detectors.
\end{abstract}

\maketitle

\section{Introduction}

Extreme-mass-ratio inspirals (EMRIs) provide one of the most sensitive probes of spacetime geometry around massive compact objects. In general relativity, an isolated stationary astrophysical black hole is expected to be described by the Kerr metric, whose external gravitational field is completely characterized by its mass and spin~\cite{Kerr1963,Carter1971,Chandrasekhar1983}. A small compact object inspiralling into such a massive central body can therefore act as a tracer of the background geometry, and the emitted gravitational waves can encode information about the multipolar structure of the central object~\cite{Ryan1995Mapping,Ryan1997}. Since an EMRI may remain in the strong-field region for a very large number of orbital cycles, even a small deviation from the Kerr geometry can accumulate into an appreciable phase shift in the gravitational waveform~\cite{Hughes2000,BarackCutler2004,Gair2013}. This makes EMRIs an important target for future space-based gravitational-wave detectors and a powerful tool for testing the Kerr hypothesis in the strong-field regime~\cite{AmaroSeoane2017LISA, Babak2017LISA}. It is therefore necessary to understand how parametrized deviations from the Kerr spacetime modify bound orbital motion and the
associated gravitational-wave phenomenology.

The orbital dynamics of test bodies in the Kerr spacetime has been extensively studied and provides the theoretical foundation for EMRI modeling. Because the Hamilton--Jacobi equation is separable, bound timelike geodesics can be characterized by the energy, the axial angular momentum, and the Carter constant, and their motion can be described in terms of three fundamental frequencies associated with the radial, polar, and azimuthal directions~\cite{Carter1968,Schmidt2002,Mino2003,DrascoHughes2004,FujitaHikida2009}. When two of these frequencies become commensurate, the orbit exhibits a resonant or periodic structure, and such resonances may affect the phase evolution and parameter estimation of the EMRI waveforms~\cite{LevinPerezGiz2008,FlanaganHinderer2012,Brink2015,Berry2016,SperiGair2021,Levati2025}. In parallel, a variety of parametrized non-Kerr spacetimes have been developed to describe possible deviations from the Kerr geometry in a theory-agnostic way, including quasi-Kerr metrics, deformed metrics admitting approximate or exact constants of motion, and general axisymmetric parametrizations~\cite{CollinsHughes2004, GlampedakisBabak2006,VigelandHughes2010,VigelandYunesStein2011, JohannsenPsaltis2011,Johannsen2013,KonoplyaRezzollaZhidenko2016, DelPiano2026}. These studies provide both the dynamical tools and the phenomenological motivation for investigating how a specific non-Kerr
deformation modifies bound orbits, resonant structures, and EMRI waveforms.

Among the parametrized non-Kerr spacetimes, the rotating KZ black hole metric provides a simple and useful phenomenological model for describing deviations from the Kerr geometry. In this spacetime, the deviation is controlled by an additional parameter $\eta$, and the Kerr solution is recovered in the limit $\eta=0$~\cite{KonoplyaZhidenko2016}. KZ and other parametrized non-Kerr geometries have been investigated through black-hole shadows, gravitational lensing, X-ray reflection spectroscopy, and gravitational-wave observations~\cite{WangChenJing2016,WangChenJing2017,Younsi2016,Nampalliwar2020,Abdikamalov2021,Yu2021,ShashankBambi2022}. However, most existing studies focus on null geodesics or electromagnetic observables, while the timelike bound-orbit dynamics directly relevant to EMRIs has received comparatively less attention. In particular, it remains important to understand how the KZ deformation modifies the mapping between orbital parameters and constants of motion, shifts the fundamental-frequency resonances of eccentric and inclined orbits, and changes the morphology of resonant trajectories and their associated gravitational-wave signals. This provides the main motivation for studying EMRI-related timelike geodesics and resonant dynamics in the KZ spacetime.

In this work, we investigate timelike bound motion, resonant periodic orbits, and their gravitational-wave signatures in the rotating KZ black hole spacetime. We parameterize eccentric and inclined bound orbits by $(p,e,z_1)$, derive the corresponding constants of motion $(E,L_z,Q)$, and use the orbital frequencies to identify resonant configurations. We then investigate a range of low-order radial-polar resonances for inclined orbits and radial-azimuthal resonances for equatorial eccentric orbits, and examine how the deformation parameter shifts the resonance locations and modifies the corresponding closed trajectories. For representative equatorial resonant orbits, we further calculate quadrupole-kludge waveforms, analyze their frequency-domain characteristics, and compare the characteristic strain with representative detector noise curves.

The rest of this paper is organized as follows. In Sec.~II, we introduce the KZ spacetime, derive the timelike geodesic equations, and discuss equatorial circular orbits and the innermost stable circular orbit (ISCO). In Sec.~III, we study generic bound orbits, resonant periodic trajectories, and their gravitational-wave signatures. Afterward, we also discuss the partial degeneracy between the KZ deformation and Kerr spin. Section~IV summarizes our main results and discusses their implications. Throughout this paper, we use geometrized units $G=c=1$ unless otherwise stated.

\section{Background}
The KZ deformed Kerr metric introduced in Ref.~\cite{KonoplyaZhidenko2016} describes a rotating black hole spacetime characterized by an additional deformation parameter $\eta$. In Boyer--Lindquist coordinates, the metric is given by
\begin{align}
ds^{2} ={}&
-\left(1-\frac{2Mr^2+\eta}{r\rho^{2}}\right)dt^{2}
+\frac{\rho^{2}}{\Delta}dr^{2}
+\rho^{2} d\theta^{2} \notag\\
&+\sin^{2}\theta\left[
r^{2}+a^{2}
+\frac{(2Mr^{2}+\eta)a^2\sin^{2}\theta}{r\rho^{2}}
\right]d\phi^{2} \notag\\
&-\frac{2(2Mr^2+\eta)a\sin^{2}\theta}{r\rho^{2}}dtd\phi .
\label{eq:KZmetric}
\end{align}
The metric functions are
\begin{equation}
\Delta=r^{2}+a^{2}-2Mr-\frac{\eta}{r},\;\;\;\;\;\;\;\;\;\; \rho^{2}=r^{2}+a^{2}\cos^{2}\theta.
\end{equation}
Here, $M$, $a$, and $\eta$ denote the black hole mass, spin parameter, and deformation parameter, respectively. The parameter $\eta$ characterizes deviations from the Kerr metric. The Kerr metric is recovered for $\eta=0$. The conditions for the existence of a black hole horizon were analyzed in Ref.~\cite{WangChenJing2016}.
For $a<M$, the negative-deformation black hole branch is bounded by $\eta_{c1}\leq\eta \leq 0$, where
\begin{align}
&\eta_{c1}\equiv-\frac{2}{27}(\sqrt{4M^{2}-3a^{2}}+2M)^{2}(\sqrt{4M^{2}-3a^{2}}-M),
\end{align}
For $a>M$, a black hole horizon exists only for $\eta>0$. In this work, we restrict our attention to the negative-deformation branch with $a<M$, namely $\eta_{c1}\leq \eta \leq 0$. For parameter values outside the corresponding black hole regions, the spacetime has no event horizon and describes a naked singularity.

As in the Kerr spacetime, the KZ metric
admits three independent constants of geodesic motion. Here,
\(E\) and \(L_z\) denote the energy and axial angular momentum
per unit rest mass, respectively, while \(Q\) denotes the Carter
constant per unit rest mass squared.

The separability of the Hamilton--Jacobi equation yields the Carter
constant and allows the radial and polar motions to be written in terms
of the effective potentials~\cite{Carter1968}
\begin{equation}
R(r)
=
\left[(r^2+a^2)E-aL_z\right]^2
-\Delta
\left[
r^2+(L_z-aE)^2+Q
\right],
\label{eq:RKZ}
\end{equation}
and
\begin{equation}
\Theta(\theta)
=
Q-
\left[
a^2(1-E^2)
+\frac{L_z^2}{\sin^2\theta}
\right]\cos^2\theta .
\label{eq:ThetaKZ}
\end{equation}

The geodesic equations of motion are then given by
\begin{align}
\rho^4 \dot{r}^{\,2} &= R(r),
\label{eq:radial_motion}
\\
\rho^4 \dot{\theta}^{\,2} &= \Theta(\theta).
\label{eq:theta_motion}
\end{align}
For the azimuthal component, one obtains
\begin{equation}
\begin{aligned}
\rho^2 \dot{\phi}
\equiv{}&
\Phi_r(r)+\Phi_\theta(\theta).
\end{aligned}
\label{eq:phi_motion}
\end{equation}
The corresponding radial and polar contributions are
\begin{align}
&\Phi_r(r)=aE\left(\frac{r^2+a^2}{\Delta}-1\right)
-\frac{a^2L_z}{\Delta},\notag\\
&\Phi_\theta(\theta)=\csc^2\theta\, L_z,\notag
\end{align}
and the temporal component is
\begin{equation}
\begin{aligned}
\rho^2 \dot{t}
\equiv{}
T_r(r)+T_\theta(\theta).
\end{aligned}
\label{eq:t_motion}
\end{equation}
Its separated contributions are
\begin{align}
&T_r(r)=E\frac{(r^2+a^2)^2}{\Delta}
+aL_z\left(1-\frac{r^2+a^2}{\Delta}\right),\notag\\
&T_\theta(\theta)=-a^2E\sin^2\theta.\notag
\end{align}
An overdot denotes differentiation with respect to the proper time $\tau$. For numerical convenience, we introduce the Mino time $\lambda$, defined by $d\lambda={d\tau}/{\rho^2}$. With this parametrization, the geodesic equations take the form
\begin{equation}
\begin{gathered}
\left(\frac{dr}{d\lambda}\right)^2=R(r),\quad
\left(\frac{d\theta}{d\lambda}\right)^2=\Theta(\theta),\\[0.8ex]
\frac{d\phi}{d\lambda}=\Phi_r(r)+\Phi_\theta(\theta),\quad
\frac{dt}{d\lambda}=T_r(r)+T_\theta(\theta).
\end{gathered}
\label{eq:mino_eqs}
\end{equation}

As a reference for bound-orbit stability, we consider equatorial circular orbits in the KZ spacetime. For equatorial motion, one has $\theta=\pi/2$ and $Q=0$. The radial potential reduces to
\begin{equation}
R(r)=\left[E(r^2+a^2)-aL_z\right]^2
-\Delta\left[r^2+(L_z-aE)^2\right],
\end{equation}
and circular orbits are determined by
\begin{equation}
R(r)=0,\qquad \frac{dR(r)}{dr}=0 .
\end{equation}
The ISCO is determined by the marginal-stability condition
\begin{equation}
\frac{d^2R(r)}{dr^2}=0 .
\end{equation}

\begin{figure*}[t]
    \centering
    \includegraphics[width=0.95\textwidth]{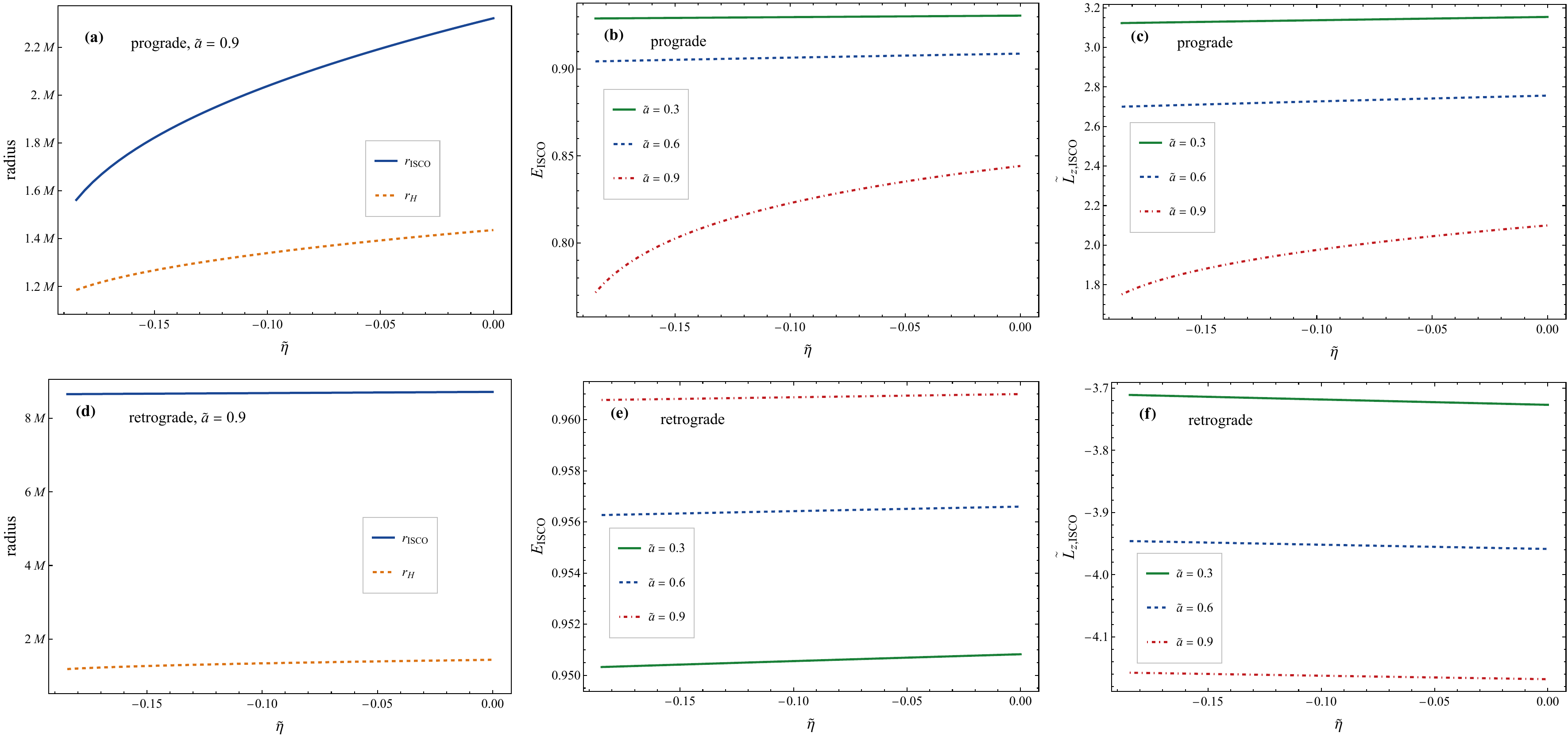}
    \caption{
    ISCO quantities for equatorial circular orbits in the KZ spacetime.
    The upper and lower rows correspond to prograde and retrograde orbits, respectively.
    Panels (a) and (d) show $r_{\rm ISCO}$ and the horizon radius $r_{\rm H}$ for $\tilde a=0.9$.
    Panels (b), (e) and (c), (f) show $E_{\rm ISCO}$ and $\tilde L_{z,\rm ISCO}$ for different spin parameters.
    }
    \label{fig:isco_quantities}
\end{figure*}
For convenience in presenting the numerical results, we introduce the dimensionless quantities
\begin{equation}\label{eq:dimensionless}
\begin{gathered}
\tilde r = \frac{r}{M}, \qquad
\tilde a = \frac{a}{M}, \qquad
\tilde\eta = \frac{\eta}{M^3},\\
\tilde L_z = \frac{L_z}{M}, \qquad
\tilde Q = \frac{Q}{M^2}.
\end{gathered}
\end{equation}
The specific energy $E$ is already dimensionless and therefore requires
no further rescaling.
FIG.~\ref{fig:isco_quantities} shows that the KZ deformation produces a systematic shift in the ISCO properties. For prograde orbits, $r_{ISCO}$, $E_{ISCO}$, and $L_{z,ISCO}$ decrease with increasing $|\eta|$. The dependence is considerably weaker for retrograde orbits, although the deformation still produces a noticeable shift in the corresponding ISCO quantities.

\section{Bound orbits, resonances, and gravitational-wave signatures}
\subsection{Parameterization of geodesic motion}
We next consider generic bound timelike geodesics. Following the standard parameterization used for bound Kerr geodesics~\cite{Schmidt2002,DrascoHughes2004,FujitaHikida2009}, the radial motion is described by the semi-latus rectum $p$ and eccentricity $e$ as
\begin{equation}
r(\chi_r)=\frac{pM}{1+e\cos\chi_r},
\label{eq:r_param}
\end{equation}
where $\chi_r$ is the radial phase angle. The radial turning points are therefore
\begin{equation}
r_{1}=\frac{pM}{1-e},\qquad
r_{2}=\frac{pM}{1+e},
\end{equation}
and as $\chi_r$ varies from 0 to $\pi$, $r$ ranges from periapsis $r_2$ to apoapsis $r_1$. Unlike in the Kerr spacetime, the radial potential $R(r)$ is not a quartic polynomial because of the $r^{-1}$ term. Instead, $rR(r)$ is a quintic polynomial and can be factorized as
\begin{equation}
rR(r)=(1-E^2)(r_1-r)(r-r_2)(r-r_3)(r-r_4)(r-r_5) ,
\label{eq:R(r)_quintic}
\end{equation}
where the roots are ordered as $r_5\leq r_4\leq r_3 < r_2 < r_1$ for the bound orbits considered here. The physical radial motion is confined to $r_2\leq r\leq r_1$, with $R(r)>0$ for $r\in \left(r_2,r_1\right)$. Here $p_3$, $p_4$, and $p_5$ are defined in terms of $r_3$, $r_4$, and $r_5$ by
\begin{equation}
r_{3}=\frac{p_3M}{1-e},\qquad
r_{4}=\frac{p_4M}{1+e},\qquad
r_{5}=\frac{p_5M}{1+e},
\end{equation}
which gives
\begin{equation}
\frac{d\chi_r}{d\lambda}=\frac{d\chi_r}{dr}\frac{dr}{d\lambda}=\frac{M\sqrt{1-E^2}}{\sqrt{p}(1-e^2)}\sqrt{\frac{\zeta \xi \varpi}{1+e}},
\label{eq:dchirdlambda}
\end{equation}
where
\begin{align}
\zeta = (p-p_3)-e(p+p_3\cos\chi_r),\notag \\
\xi = (p-p_4)+e(p-p_4\cos\chi_r),\notag \\
\varpi = (p-p_5)+e(p-p_5\cos\chi_r),\notag
\label{chirlambdapara}
\end{align}
The auxiliary parameters $p_3$, $p_4$, $p_5$ are introduced only to cast Eq.~(\ref{eq:dchirdlambda}) into a compact form, and the radial dynamics is determined by the roots $r_3$, $r_4$, $r_5$.

For the $\theta$ component, defining $z=\cos\theta$, we follow the
convention of Refs.~\cite{vanDeMeent2020,Drummond2023} and rewrite
$\Theta(\theta)$ in Eq.~\eqref{eq:ThetaKZ} as
\begin{equation}
\Theta(\theta)
=
\frac{z_1^2-z^2}{1-z^2}
\left[
z_2^2-a^2(1-E^2)z^2
\right],
\label{eq:theta_factorized}
\end{equation}
where
\begin{equation}
z_2^2
=
a^2(1-E^2)
+
\frac{L_z^2}{1-z_1^2}.
\label{eq:z2_dimensional}
\end{equation}
Here $z_1=\cos\theta_{\rm min}$, with $0\leq z_1\leq1$, denotes the
physical polar turning point. The quantity $z_2$ follows the convention
of Refs.~\cite{vanDeMeent2020,Drummond2023} and is a rescaled parameter
associated with the second zero of the polar potential. This convention
is particularly convenient because it remains regular in the
$a\to0$ limit.
The polar motion can then be parameterized as
\begin{equation}
\cos\theta=z_1\cos\chi_\theta ,
\label{eq:theta_param}
\end{equation}
where $\chi_\theta$ is the polar phase angle. Substituting this
parameterization into the polar geodesic equation gives
\begin{equation}
\frac{d\chi_\theta}{d\lambda}
=
\sqrt{
z_2^2-a^2(1-E^2)z_1^2\cos^2\chi_\theta
},
\label{eq:chitheta}
\end{equation}
which determines the evolution of the polar phase.

\subsection{Mapping between orbital parameters and constants of motion}

For numerical calculations, we determine the dimensionless constants of motion 
\((E,\tilde L_z,\tilde Q)\) from the orbital parameters \((p,e,z_1)\). In this subsection, we focus on eccentric bound orbits with \(e>0\). Using the dimensionless quantities \eqref{eq:dimensionless} introduced in Sec.~II, the polar turning-point condition gives
\begin{equation}
\tilde Q
=
z_1^2
\left[
\tilde a^2(1-E^2)
+
\frac{\tilde L_z^2}{1-z_1^2}
\right].
\label{eq:Q_from_z1}
\end{equation}
Substituting Eq.~\eqref{eq:Q_from_z1} into the radial potential, the
dimensionless KZ radial potential,
\(\tilde R\equiv R/M^4\), can be written as a quadratic form in
\(E\) and \(\tilde L_z\):
\begin{equation}
\tilde R(\tilde r)
=
\tilde f(\tilde r)E^2
-2\tilde g(\tilde r)E\tilde L_z
-\tilde h(\tilde r)\tilde L_z^2
-\tilde d(\tilde r),
\label{eq:R_quadratic_ELQ}
\end{equation}
where
\begin{align}
\tilde f(\tilde r)
={}&
\tilde r^4
+
\tilde a^2
\left[
\tilde r(\tilde r+2)
+
z_1^2\tilde\Delta
+
\frac{\tilde\eta}{\tilde r}
\right],
\label{eq:fKZ}
\\
\tilde g(\tilde r)
={}&
\tilde a
\left(
2\tilde r
+
\frac{\tilde\eta}{\tilde r}
\right),
\label{eq:gKZ}
\\
\tilde h(\tilde r)
={}&
\tilde r(\tilde r-2)
+
\frac{z_1^2}{1-z_1^2}\tilde\Delta
-
\frac{\tilde\eta}{\tilde r},
\label{eq:hKZ}
\\
\tilde d(\tilde r)
={}&
\left(
\tilde r^2+\tilde a^2z_1^2
\right)
\tilde\Delta .
\label{eq:dKZ}
\end{align}
Here
\begin{equation}
\tilde\Delta
=
\tilde r^2+\tilde a^2-2\tilde r
-\frac{\tilde\eta}{\tilde r}.
\end{equation}
In the limit \(\tilde\eta=0\), these functions reduce to their Kerr
counterparts.

At the two radial turning points, one has
\(\tilde R(\tilde r_1)=\tilde R(\tilde r_2)=0\). We denote
\begin{equation}
\tilde f_i=\tilde f(\tilde r_i),\quad
\tilde g_i=\tilde g(\tilde r_i),\quad
\tilde h_i=\tilde h(\tilde r_i),\quad
\tilde d_i=\tilde d(\tilde r_i),
\end{equation}
where \(i=1,2\). The two turning-point equations are therefore
\begin{align}
\tilde f_1E^2
-2\tilde g_1E\tilde L_z
-\tilde h_1\tilde L_z^2
-\tilde d_1
&=0,
\\
\tilde f_2E^2
-2\tilde g_2E\tilde L_z
-\tilde h_2\tilde L_z^2
-\tilde d_2
&=0.
\end{align}

Following the standard algebraic elimination procedure, we introduce
\begin{align}
\kappa
&\equiv
\tilde d_1\tilde h_2-\tilde d_2\tilde h_1,
&
\epsilon
&\equiv
\tilde d_1\tilde g_2-\tilde d_2\tilde g_1,
\notag\\
\varrho
&\equiv
\tilde f_1\tilde h_2-\tilde f_2\tilde h_1,
&
\delta
&\equiv
\tilde f_1\tilde g_2-\tilde f_2\tilde g_1,
\notag\\
\sigma
&\equiv
\tilde g_1\tilde h_2-\tilde g_2\tilde h_1 .
\label{eq:determinants_mapping}
\end{align}

To distinguish the prograde and retrograde branches, we define the orbit-sense parameter
\begin{equation}
D\equiv \operatorname{sgn}(\tilde L_z),
\notag
\end{equation}
where \(D=+1\) corresponds to prograde motion and \(D=-1\) corresponds to retrograde motion for \(\tilde a>0\). Adapting Schmidt's algebraic construction, the energy is given by
\begin{equation}
E^2 = \frac{ \kappa\varrho+2\epsilon\sigma
-2D\sqrt{\sigma\left(\sigma\epsilon^2+\varrho\epsilon\kappa-\delta\kappa^2
\right)}}{\varrho^2+4\delta\sigma}.
\label{eq:E2_mapping_with_D}
\end{equation}

Once \(E\) is obtained, the dimensionless axial angular momentum is
determined by
\begin{equation}
\tilde L_z
=
-\frac{\tilde g_iE}{\tilde h_i}
+
D\sqrt{
\frac{\tilde g_i^2E^2}{\tilde h_i^2}
+
\frac{\tilde f_iE^2-\tilde d_i}{\tilde h_i}
},
\label{eq:Lz_mapping_with_D}
\end{equation}
where \(i=1\) and \(2\) correspond to the apoapsis
\(\tilde r_1\) and periapsis \(\tilde r_2\), respectively.
Since both turning points satisfy
\(\tilde R(\tilde r_i)=0\), either turning point can be used to
evaluate \(\tilde L_z\), and the two choices agree up to numerical
precision. The physical branch is selected by imposing the value of orbit-sense parameter $D$.

The dimensionless Carter constant is then obtained from
Eq.~\eqref{eq:Q_from_z1}, while $z_2$ follows from
Eq.~\eqref{eq:z2_dimensional}.

\subsection{Resonant trajectories and gravitational-wave signatures}
\label{subsec:resonant_waveforms}

Bound geodesic motion in the KZ spacetime is characterized by the
radial, polar, and azimuthal fundamental frequencies. In terms of the
Mino time, the radial and polar periods are
\begin{equation}
\Lambda_r=\int_0^{2\pi}\frac{d\chi_r}{F_r(\chi_r)},
\qquad
\Lambda_\theta=\int_0^{2\pi}\frac{d\chi_\theta}{F_\theta(\chi_\theta)},
\label{eq:mino_periods}
\end{equation}
where $F_r=d\chi_r/d\lambda$ and
$F_\theta=d\chi_\theta/d\lambda$ are given by
Eqs.~\eqref{eq:dchirdlambda} and \eqref{eq:chitheta}. The corresponding
Mino-time frequencies are
\begin{equation}
\Upsilon_r=\frac{2\pi}{\Lambda_r},
\qquad
\Upsilon_\theta=\frac{2\pi}{\Lambda_\theta}.
\end{equation}
We further define
\begin{equation}
\Upsilon_\phi=\left\langle\frac{d\phi}{d\lambda}\right\rangle,
\qquad
\Gamma=\left\langle\frac{dt}{d\lambda}\right\rangle,
\end{equation}
where the brackets denote averages over the bound orbital motion. The coordinate-time frequencies are
\begin{equation}
\Omega_r=\frac{\Upsilon_r}{\Gamma},
\qquad
\Omega_\theta=\frac{\Upsilon_\theta}{\Gamma},
\qquad
\Omega_\phi=\frac{\Upsilon_\phi}{\Gamma}.
\label{eq:coordinate_frequencies}
\end{equation}

For generic non-equatorial orbits, the radial--polar resonance is
specified by
\begin{equation}
\frac{\Omega_\theta}{\Omega_r}
=\frac{\Upsilon_\theta}{\Upsilon_r}
=\frac{m}{n}.
\label{eq:theta_r_resonance}
\end{equation}
For equatorial eccentric orbits, $\theta=\pi/2$ and $Q=0$, and the
radial--azimuthal resonance is defined by
\begin{equation}
\frac{\Omega_\phi}{\Omega_r}
=\frac{\Delta\phi}{2\pi}
=\frac{m}{n},
\qquad
\Delta\phi=\int_0^{2\pi}
\frac{d\phi/d\lambda}{d\chi_r/d\lambda}\,d\chi_r .
\label{eq:r_phi_resonance}
\end{equation}
Here, resonance refers to an exact commensurability of the fundamental frequencies of bound geodesics, rather than a transient resonance crossing during a radiation-reaction-driven inspiral.

For a radial--polar resonance, the radial and polar motions return
simultaneously to their initial values after $n$ radial cycles and
$m$ polar cycles, yielding a closed trajectory in the meridional-plane
projection. For an equatorial radial--azimuthal resonance, the spatial
orbit closes after $n$ radial cycles and $m$ azimuthal cycles.

\begin{figure}[t]
\centering
\begin{minipage}{0.48\textwidth}
\centering
\includegraphics[width=\linewidth]{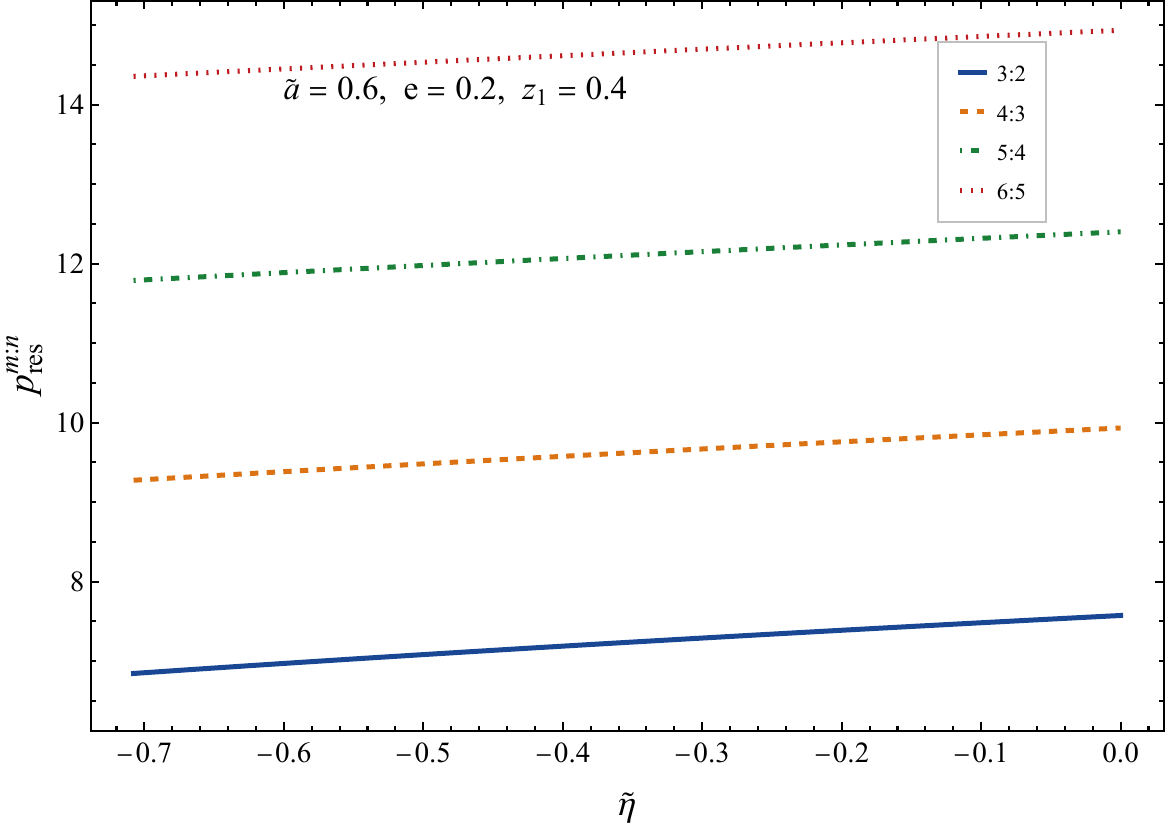}
\end{minipage}
\hfill
\begin{minipage}{0.48\textwidth}
\centering
\includegraphics[width=\linewidth]{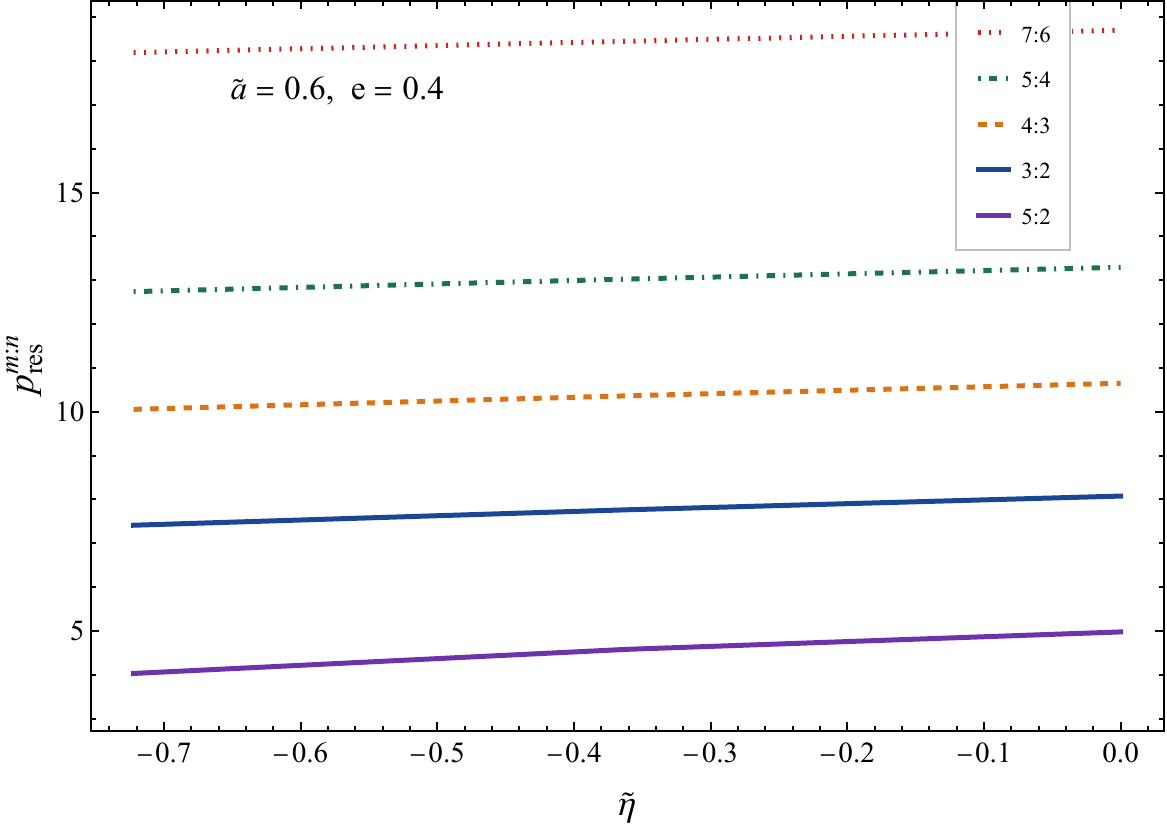}
\end{minipage}
\caption{Resonant semi-latus rectum $p_{\rm res}$ as a function of the
KZ deformation parameter $\tilde \eta$. The first panel shows non-equatorial
radial--polar resonances for $\tilde a=0.6$, $e=0.2$, and $z_1=0.4$. The
second panel shows equatorial radial--azimuthal resonances for $\tilde a=0.6$
and $e=0.4$.}
\label{fig:resonance_pres_eta}
\end{figure}

FIG.~\ref{fig:resonance_pres_eta} shows the resonant semi-latus rectum
$p_{\rm res}$ as a function of $\tilde \eta$. For the radial--polar resonances,
we set $\tilde a=0.6$, $e=0.2$, and $z_1=0.4$, while the equatorial
radial--azimuthal resonances are calculated for $\tilde a=0.6$ and $e=0.4$.
The negative deformation shifts the resonances inward.

\begin{figure*}[t]
\centering
\includegraphics[width=0.95\textwidth]{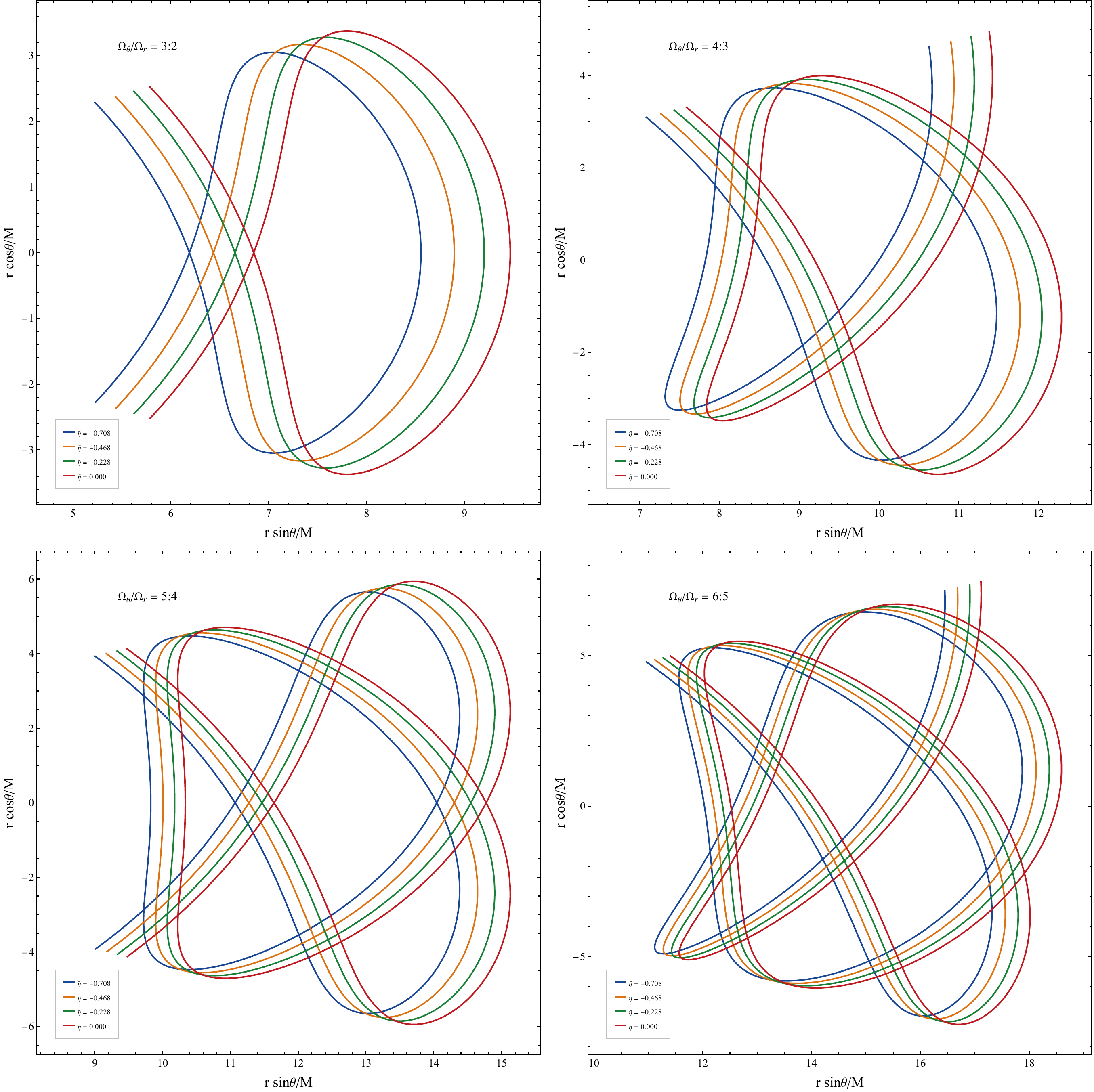}
\caption{Meridional-plane projections of representative radial--polar resonant trajectories. Each panel corresponds to a different value of $\Omega_\theta/\Omega_r=m/n$, and the curves denote selected values of
$\tilde \eta$.}
\label{fig:theta_r_meridian_orbits}
\end{figure*}

The meridional-plane projection of a radial--polar resonant orbit is
defined by
\begin{equation}
R_\perp=r\sin\theta,
\qquad
Z=r\cos\theta.
\end{equation}
As shown in FIG.~\ref{fig:theta_r_meridian_orbits}, different resonance ratios produce distinct orbital patterns in the meridional plane, while the KZ deformation changes their spatial scale through the shift of $p_{\rm res}$.

Following the numerical-kludge prescription~\cite{Ruffini1981,Gair2005,Babak2007}, we identify the Boyer--Lindquist coordinates of the geodesic trajectory with spherical polar coordinates in an auxiliary flat spacetime and construct the corresponding Cartesian trajectory. For equatorial periodic motion, this gives
\begin{equation}
x=r\cos\phi,
\qquad
y=r\sin\phi.
\end{equation}

The waveforms are calculated as a post-processing of the prescribed geodesic trajectories, without including the backreaction of the emitted radiation on the orbital motion. Thus, the orbital parameters and constants of motion remain fixed during the waveform calculation. The gravitational waveform is evaluated using the leading-order flat-spacetime quadrupole formula~\cite{Thorne1980}. In this framework, the source is
characterized by the mass quadrupole moment
\begin{align}
&I_{ij}(t_r)
=\int x_i'x_j' T_{00}(t_r,\bm x')\,d^3x',
\label{eq:mass_quadrupole_integral}
\\
\text{with}~~&T_{00}(t_r,\bm x')
=\mu\,\delta^{(3)}\!\left[\bm x'-\bm Z(t_r)\right],
\label{eq:point_particle_T00}
\end{align}
where 
\(x_i'\) are the Cartesian source coordinates, and \(T_{00}\) is the
energy-density component of the stress-energy tensor of the particle.

Here \(t_r=t-D_L/c\) is the retarded time. The far-zone metric perturbation is
\begin{equation}
h_{ij}(t_r)
=\frac{2G}{c^4D_L}
\frac{d^2 I_{ij}(t_r)}{dt_r^2}.
\label{eq:quadrupole_formula}
\end{equation}

For an observer in the direction
\begin{equation}
\bm n=
(\sin\Theta\cos\Phi,\,
 \sin\Theta\sin\Phi,\,
 \cos\Theta),
\end{equation}
the transverse basis vectors are
\begin{align}
\bm e_\Theta
&=(\cos\Theta\cos\Phi,\,
   \cos\Theta\sin\Phi,\,
   -\sin\Theta),
\notag\\
\bm e_\Phi
&=(-\sin\Phi,\,\cos\Phi,\,0).
\label{eq:polarization_basis}
\end{align}

\begin{figure*}[t]
\centering
\includegraphics[width=0.98\textwidth]{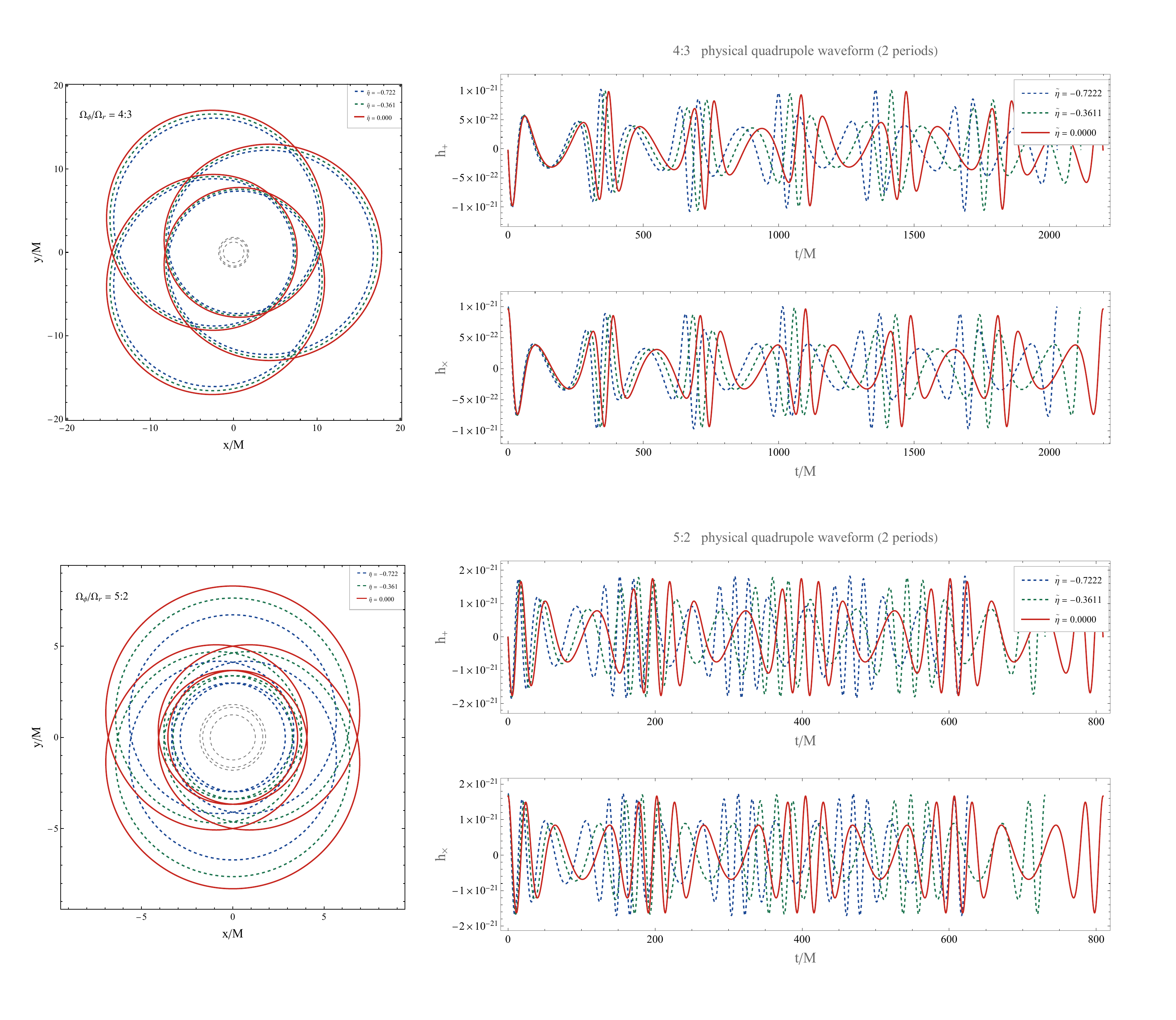}
\caption{Equatorial periodic trajectories and the corresponding quadrupole waveforms. The left and right columns show the trajectories and two waveform periods, respectively. The upper and lower rows correspond to the $4{:}3$ and $5{:}2$ resonances. The blue dashed, green dashed, and red solid curves denote $\tilde \eta=\tilde \eta_{c1}$, $\tilde \eta=\tilde \eta_{c1}/2$, and $\tilde \eta=0$, respectively. The dashed circles represent the event horizons.}
\label{fig:rphi_orbits_waveforms}
\end{figure*}

The projected components are
\begin{align}
h_{\Theta\Theta}={}&\cos^2\Theta\left( h_{xx}\cos^2\Phi
 +h_{xy}\sin 2\Phi +h_{yy}\sin^2\Phi\right) \notag\\
&+h_{zz}\sin^2\Theta-\sin 2\Theta
\left( h_{xz}\cos\Phi+h_{yz}\sin\Phi\right),
\label{eq:h_thetatheta}
\end{align}
\begin{align}
h_{\Theta\Phi}={}&\cos\Theta\left( -\frac{1}{2}h_{xx}\sin 2\Phi
 +h_{xy}\cos 2\Phi +\frac{1}{2}h_{yy}\sin 2\Phi\right)
\notag\\
&+\sin\Theta\left( h_{xz}\sin\Phi-h_{yz}\cos\Phi\right),
\label{eq:h_thetaphi}
\end{align}
\begin{align}
h_{\Phi\Phi}={}&h_{xx}\sin^2\Phi-h_{xy}\sin 2\Phi+h_{yy}\cos^2\Phi.
\label{eq:h_phiphi}
\end{align}
The two polarizations are therefore
\begin{align}
h_+ &=\frac{1}{2}\left(h_{\Theta\Theta}-h_{\Phi\Phi}\right), \notag\\
h_\times &=h_{\Theta\Phi}.
\label{eq:wave_polarizations}
\end{align}

For the equatorial orbits considered here,
$\bm Z=(x,y,0)$ and hence $h_{xz}=h_{yz}=h_{zz}=0$.

We adopt the following source and observer parameters:
\begin{align}
&M=10^6M_\odot,\qquad\quad
\mu=10M_\odot,
\notag\\
D_L&=200\,{\rm Mpc},\qquad\quad
\Theta=\Phi=\frac{\pi}{4}.\notag
\label{eq:waveform_parameters}
\end{align}

FIG.~\ref{fig:rphi_orbits_waveforms} shows the $4{:}3$ and $5{:}2$
periodic trajectories and their waveforms. For each resonance, we consider
$\tilde\eta=\tilde\eta_{c1}$, $\tilde\eta=\tilde\eta_{c1}/2$, and $\tilde\eta=0$, with
$p_{\rm res}$ determined separately. The $5{:}2$ orbit displays a more
pronounced zoom--whirl structure and a correspondingly more intricate
waveform~\cite{GlampedakisKennefick2002,LevinPerezGiz2008,
LiKuangSang2024,LiKuang2026,YangBaiLiHan2026}. Varying $\eta$ modifies the orbital scale and period, thereby changing
the waveform phase and amplitude.

A closed equatorial resonant orbit has the coordinate-time period
\begin{equation}
T_{m:n}
=
\frac{2\pi n}{\Omega_r}
=
\frac{2\pi m}{\Omega_\phi}.
\label{eq:periodic_time}
\end{equation}

Accordingly, each waveform polarization admits the discrete harmonic
decomposition
\begin{equation}
h_A(t)
=
\sum_{k=-\infty}^{\infty}
H_{A,k}
\exp\!\left(
\frac{2\pi i k t}{T_{m:n}}
\right),
\qquad
A=+,\times,
\label{eq:harmonic_decomposition}
\end{equation}
with harmonic frequencies \(f_k=k/T_{m:n}\). Over the finite
observation interval
\(
T_{\rm obs}=20T_{m:n}
\),
the Fourier amplitudes are evaluated as
\begin{equation}
\widetilde h_A(f)
=
\int_{0}^{T_{\rm obs}}
h_A(t)e^{-2\pi i f t}\,dt.
\label{eq:htilde_def}
\end{equation}
The discrete Fourier amplitudes are fitted as functions of frequency
to characterize their spectral distribution.

\begin{figure*}[!tp]
\centering
\begin{minipage}{0.46\textwidth}
\centering
\includegraphics[width=\linewidth]{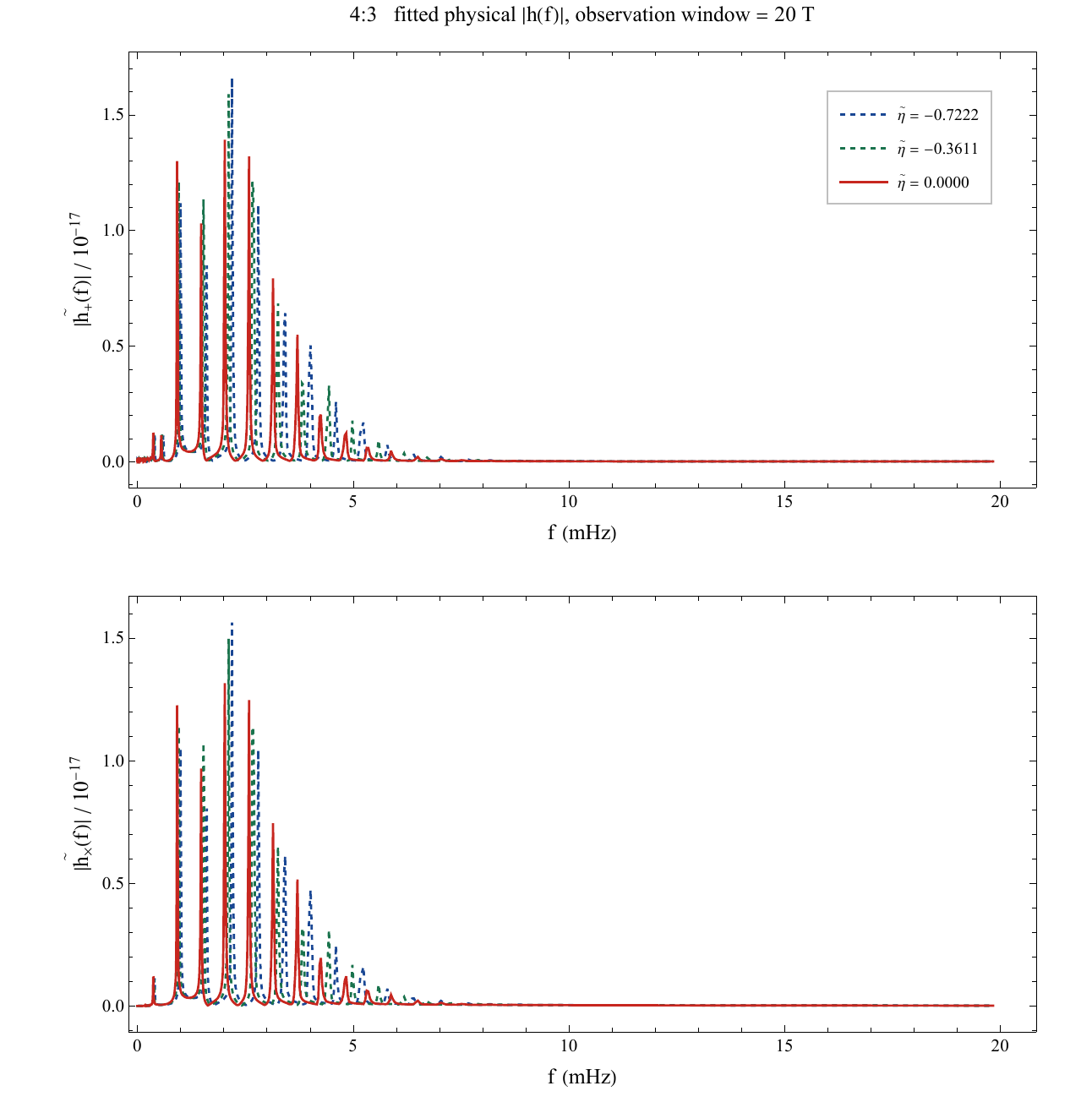}
\end{minipage}
\hfill
\begin{minipage}{0.46\textwidth}
\centering
\includegraphics[width=\linewidth]{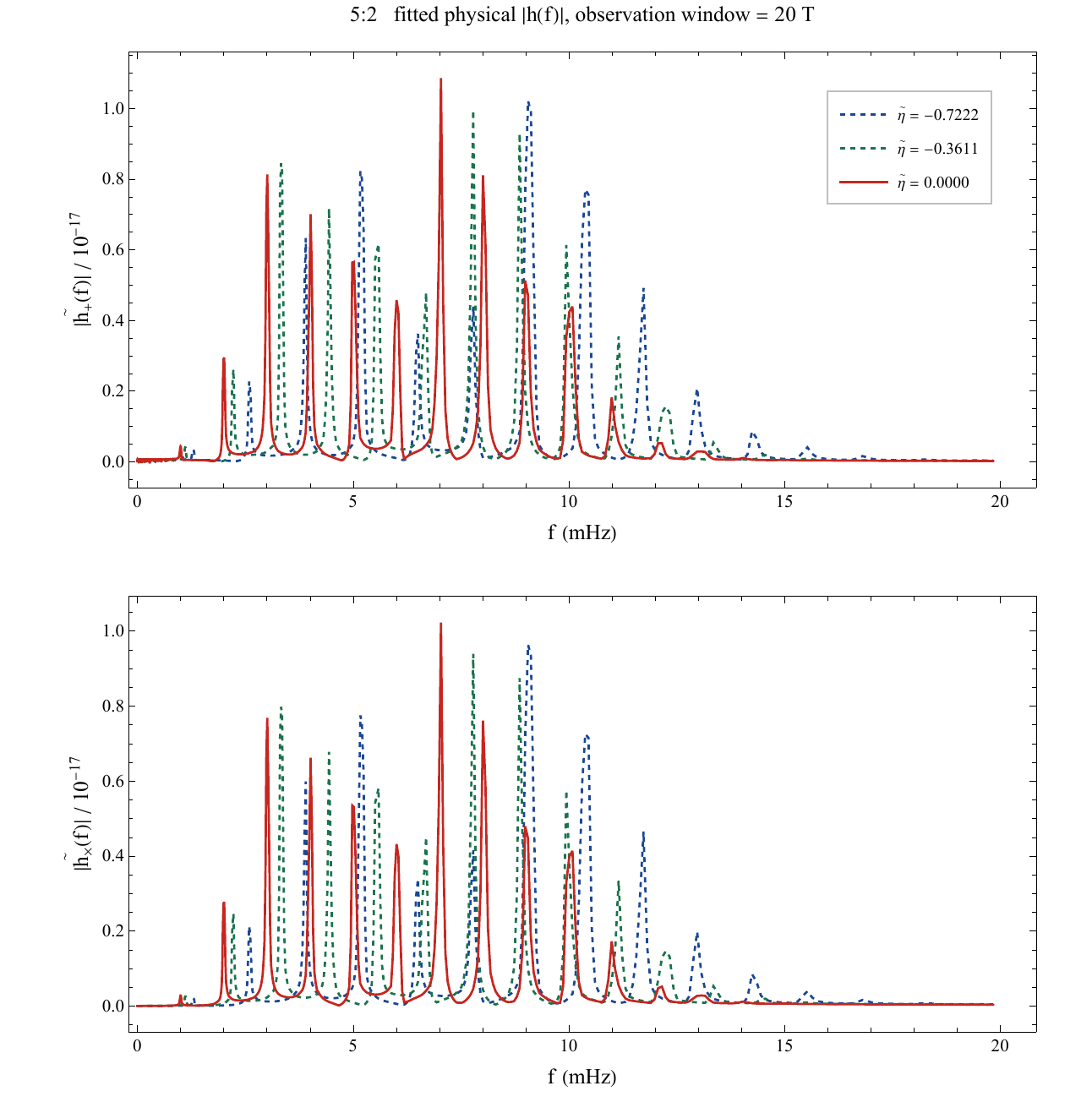}
\end{minipage}
\caption{Fitted Fourier spectra of the \(4{:}3\) (left) and \(5{:}2\)
(right) periodic-orbit waveforms obtained over the observation interval
\(T_{\rm obs}=20T_{m:n}\). The upper and lower panels show
\(\left|\widetilde h_+(f)\right|\) and
\(\left|\widetilde h_\times(f)\right|\), respectively. The blue dashed,
green dashed, and red solid curves correspond to
\(\tilde\eta=\tilde\eta_{c1}\), \(\tilde\eta=\tilde\eta_{c1}/2\), and \(\tilde\eta=0\),
respectively.}
\label{fig:rphi_fourier_spectra}

\par\vspace{1.2em}

\centering
\includegraphics[width=0.76\textwidth]
{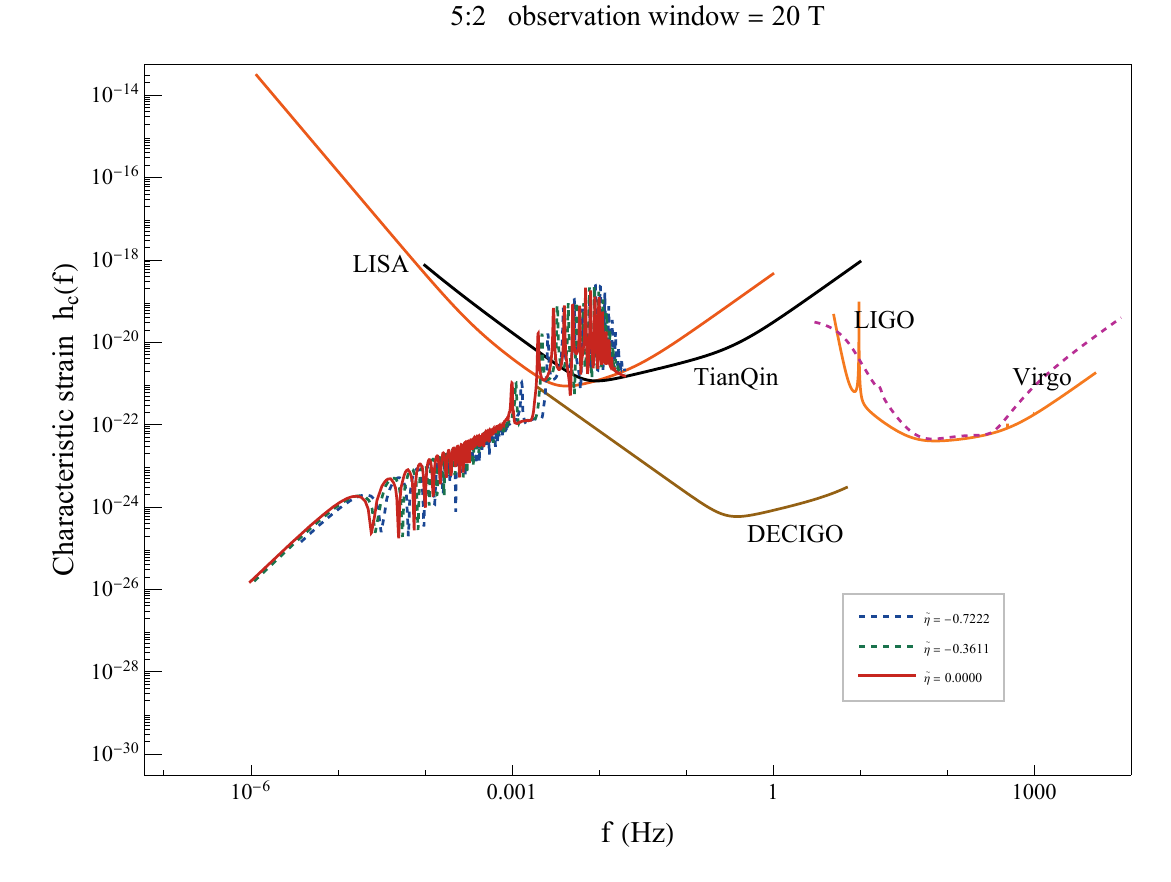}
\caption{Characteristic strain of the equatorial \(5{:}2\) periodic
orbit over the observation interval \(T_{\rm obs}=20T_{5:2}\),
together with the characteristic-noise curves of LISA, TianQin,
DECIGO, Advanced LIGO, and Advanced Virgo. The blue dashed, green
dashed, and red solid curves correspond to \(\tilde\eta=\tilde\eta_{c1}\),
\(\tilde\eta=\tilde\eta_{c1}/2\), and \(\tilde\eta=0\), respectively.}
\label{fig:rphi_characteristic_strain}
\end{figure*}

FIG.~\ref{fig:rphi_fourier_spectra} shows the fitted spectra of the \(4{:}3\) and \(5{:}2\) resonances. The \(4{:}3\) waveform is dominated by a smaller number of low-order harmonic components, whereas the zoom--whirl motion of the \(5{:}2\) orbit produces a broader spectral distribution. The KZ deformation changes both the locations and the relative amplitudes of the dominant components.

To illustrate the frequency range relevant to gravitational-wave detectors, we consider the 5:2 resonance as a representative example and calculate its characteristic strain, defined as~\cite{MooreColeBerry2015}
\begin{equation}
h_c(f)=2f
\sqrt{
\left|\widetilde h_+(f)\right|^2
+
\left|\widetilde h_\times(f)\right|^2
}.
\label{eq:characteristic_strain}
\end{equation}
The same observation interval,
\(T_{\rm obs}=20T_{5:2}\), is used for comparison with representative
detector characteristic-noise curves, which are taken from
Refs.~\cite{RobsonCornishLiu2019,LuoTianQin2016,
KawamuraDECIGO2021,AasiAdvancedLIGO2015,
AcerneseAdvancedVirgo2015}.

The results are present in FIG.~\ref{fig:rphi_characteristic_strain}, we see that the dominant components of the \(5{:}2\) signal lie in the millihertz band, while the KZ deformation modifies their frequencies and amplitudes. For the adopted source parameters and twenty-period observation windows, several dominant peaks lie above the plotted characteristic-noise curves of LISA and TianQin. The comparison in FIG.~\ref{fig:rphi_characteristic_strain} identifies the relevant detector frequency range for the adopted source parameters, but does not constitute a signal-to-noise or detectability forecast.

\subsection{Parameter degeneracy}
\label{subsec:parameter_degeneracy}

The preceding results show that a negative KZ deformation shifts the
resonances inward at fixed spin. Since the spin also enters the resonance
condition, a change in spin may compensate for this shift. We now examine
whether different combinations of spin and deformation can reproduce both
the radial extent and the orbital time scale of a given Kerr resonant
orbit.

We consider prograde equatorial orbits at fixed mass, $e=0.4$, $z_1=0$,
and $\Omega_\phi/\Omega_r=4/3$. At fixed mass and eccentricity, equal
values of $p$ give the same coordinate periapsis and apoapsis. The
resonance ratio fixes the azimuthal advance per radial cycle through
Eq.~\eqref{eq:r_phi_resonance}. These conditions alone do not fix the
orbital period across different spacetime geometries. We therefore also
compare $\Omega_r$, which determines the coordinate-time period of the
closed orbit through Eq.~\eqref{eq:periodic_time}. Because the resonance
ratio is held fixed, matching $\Omega_r$ also matches $\Omega_\phi$.

\begin{figure}[!htbp]
\centering
\includegraphics[width=\columnwidth]{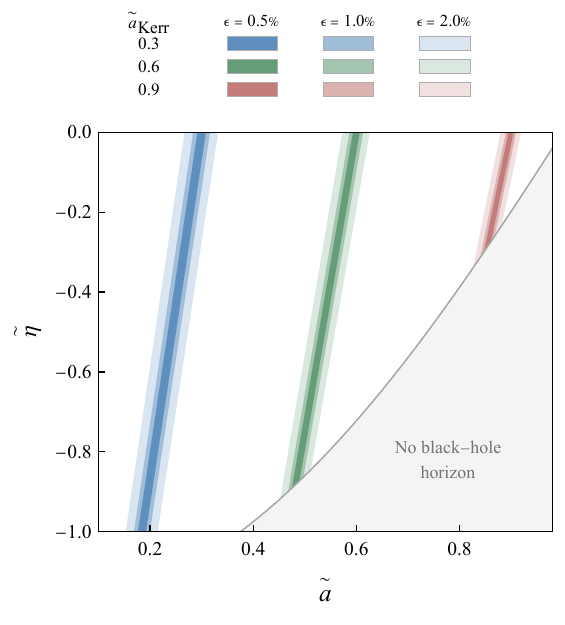}
\caption{Partial spin–deformation degeneracy for prograde equatorial 4:3 resonances at fixed mass and $e=0.4$. Blue, green, and red bands correspond to Kerr reference spins $\tilde a_{{\rm Kerr},j}=0.3,0.6,0.9$, respectively. Darker to lighter shades indicate $\varepsilon=0.5\%,1\%,2\%$, satisfying both $\delta_p^{(j)}\leq\varepsilon$ and
$\delta_\Omega^{(j)}\leq\varepsilon$, as defined in
Eqs.~\eqref{eq:degeneracy_p_difference} and \eqref{eq:degeneracy_frequency_difference}. The gray region has no black hole horizon. The bands represent
the prescribed orbital tolerances, not detector confidence regions.}
\label{fig:parameter_degeneracy}
\end{figure}

We choose three Kerr reference orbits with $\tilde\eta=0$ and
$\tilde a_{{\rm Kerr},j}=0.3,0.6,0.9$. Their resonant semi-latus recta
and dimensionless radial frequencies are denoted by $p_{{\rm K},j}$ and
$\widehat\Omega_{r,{{\rm K},j}}$, where $\widehat\Omega_r=M\Omega_r$.
For each KZ parameter pair $(\tilde a,\tilde\eta)$, we solve
Eq.~\eqref{eq:r_phi_resonance} for $p_{\rm res}$ on the prograde bound
branch outside the horizon. The constants of motion are obtained from
Eqs.~\eqref{eq:E2_mapping_with_D} and \eqref{eq:Lz_mapping_with_D}, and
the frequencies are evaluated on the same resonant orbit using
Eq.~\eqref{eq:coordinate_frequencies}. Thus, $p_{\rm res}$ and
$\widehat\Omega_r$ are calculated for each spacetime rather than varied
independently. We define their relative differences from the Kerr
reference values as
\begin{equation}
\delta_p^{(j)}=
\frac{|p_{\rm res}(\tilde a,\tilde\eta)-p_{{\rm K},j}|}
{p_{{\rm K},j}},
\label{eq:degeneracy_p_difference}
\end{equation}
and
\begin{equation}
\delta_\Omega^{(j)}=
\frac{|\widehat\Omega_r(\tilde a,\tilde\eta)-\widehat\Omega_{r,{{\rm K},j}}|}
{\widehat\Omega_{r,{{\rm K},j}}}.
\label{eq:degeneracy_frequency_difference}
\end{equation}

FIG.~\ref{fig:parameter_degeneracy} shows the regions satisfying both
$\delta_p^{(j)}\leq\varepsilon$ and
$\delta_\Omega^{(j)}\leq\varepsilon$ for
$\varepsilon=0.5\%,1\%,2\%$. These joint conditions match the coordinate
radial extent and orbital frequencies within the prescribed tolerances.
At the fixed resonance ratio, the relative differences of $\Omega_r$
and $\Omega_\phi$ are equal, so a separate azimuthal-frequency condition
is unnecessary.

It is obvious that a more negative deformation can be accompanied by a lower prograde spin while keeping both quantities close to their Kerr reference values. The finite bands illustrate partial degeneracy between spin and deformation at the chosen tolerances.

This comparison uses the coordinate-dependent parameter $p$ and
illustrative tolerances, which do not represent measurement uncertainties.
Matching the radial turning points and fundamental frequencies does not
establish agreement of the full trajectories or waveform harmonic
amplitudes and phases. Assessing observational distinguishability
requires waveform comparisons and parameter estimation.

\section{Conclusions and Discussion}

We investigated timelike bound motion, resonant periodic trajectories, and their gravitational-wave signatures in the rotating KZ black hole spacetime. Using the separability of the Hamilton--Jacobi equation, we constructed the mapping between the orbital parameters $(p,e,z_1)$ and the constants of motion $(E,L_z,Q)$, and used the orbital frequencies to identify radial-polar and equatorial radial-azimuthal resonances. The KZ deformation shifts the ISCO quantities and resonant semi-latus recta and modifies the morphology of the corresponding closed trajectories.

To characterize the gravitational-wave signatures of equatorial resonant motion, we employed the quadrupole-kludge approximation. For representative $4{:}3$ and $5{:}2$ resonances, we calculated physically scaled waveforms and analyzed their frequency-domain characteristics. The 5:2 orbit exhibits a more pronounced zoom–whirl structure and a richer harmonic content than the 4:3 orbit. Varying $\eta$ changes the orbital period, waveform phase and amplitude, and the relative strengths of the discrete Fourier components.

The present analysis characterizes how the KZ deformation affects resonant orbital dynamics and their gravitational-wave signatures. It is nevertheless based on geodesic motion and the leading-order quadrupole-kludge approximation over a finite observation interval. A realistic assessment of the observational distinguishability of the KZ deformation will require radiation-reaction-driven inspirals, more accurate relativistic waveform models, detector response, and systematic parameter-estimation studies.

The present one-parameter KZ geometry is a restricted representative of the more general separable parametrized class~\cite{KonoplyaStuchlikZhidenko2018}. More general parametrizations may involve additional deformation parameters that modify the radial and polar fundamental frequencies, potentially shifting the resonance locations and altering the corresponding orbital structures. The few-parameter description of strong-field observables~\cite{KonoplyaZhidenko2020} provides further motivation for extending the present analysis to include these additional deformation parameters in future work.

\acknowledgments
We would like to thank Dr. Yong-Zhuang Li for his useful discussions and valuable suggestions on this work. This work was partially supported by the National Natural Science Foundation of
China under Grants No. 12375046 and No. 12375054.

\FloatBarrier

\end{document}